# Precise electronic structures of amorphous solids: unraveling the color origin and photocatalysis of black titania

Jing Zhong [1,2,3], Zhengtao Xu [5], Jian Lu [1,3,4,6,*], Yang Yang Li [1,2,3,4,6,*]

[1] Hong Kong Branch of National Precious Metals Material Engineering Research Centre, City University of Hong Kong, Hong Kong, China

[2] Center of Super-Diamond and Advanced Films (COSDAF), City University of Hong Kong, Hong Kong, China

[3] Department of Materials Science and Engineering, City University of Hong Kong, Hong Kong, China

[4] Department of Mechanical Engineering, Greater Bay Joint Division, Shenyang National Laboratory for Materials Science, City University of Hong Kong, Hong Kong, China

[5] Department of Chemistry, City University of Hong Kong, Hong Kong, China

[6] Centre for Advanced Structural Materials, City University of Hong Kong Shenzhen Research Institute, Greater Bay Joint Division, Shenyang National Laboratory for Materials Science, 8 Yuexing 1st Road, Shenzhen Hi-Tech Industrial Park, Nanshan District, Shenzhen, China

[*] Email: jianlu@cityu.edu.hk; yangli@cityu.edu.hk

**Abstract**

Water splitting through efficient catalysts represents an ultimate solution for carbon neutrality within 40 years. To achieve this goal, amorphous photocatalysts are noted for their promising performances. Among them, the best known is black titania (amorphous $TiO_x$, $x \leq 2$). However, despite the large number of studies on black titania, its color origin, structure-property relationship, and photocatalytic mechanism remain a topic of hot debate, largely due to the difficulty to calculate its precise electronic structure. Here, using *ab initio* molecular dynamics simulations, we report the precise electronic structures of black titania and further reveal the generic evolution pattern of the electronic structures of covalent compounds upon amorphization and reduction. Moreover, surface adsorption of the co-catalyst atoms (e.g., Pt) on amorphous substances is simulated for the first time: the disordered surface enables easy accommodation of the co-catalyst atoms, while the extended electronic states facilitate the separation of photo-induced electrons and holes, beneficial for hydrogen evolution. This study elucidates the workings of amorphous catalysts and offers practical guidance for enhancing their performances.

**Introduction**

Amorphous solid catalysts are noted to outperform their crystalline counterparts[1,2] in several reactions, e.g., for water splitting[3]. The higher performance is generally ascribed to the reduced bandgaps or more reactive surfaces[4,5]. However, little is known about the underlying structure-property relationship, as it is difficult to calculate the precise electronic structures of the nonperiodic, amorphous structure. The challenge lies in how to simulate the disordered atomic positions accurately enough for computing the electronic structures. One common method is the melt melt-quench quench approach, which, through *ab initio* molecular dynamics (AIMD) simulations should guarantee accuracy of the resultant amorphous structures. But AIMD simulations can be prohibitively costly and too slow[6], especially for systems involving hundreds of atoms. Encouragingly, using second-generation Car-Parrinello molecular dynamics (SGCPMD) simulations[7-9], AIMD was recently applied to amorphous germanium antimony telluride and successfully simulated its crystallization process[10]. This SGCPMD approach combines the advantages of both Born-Oppenheimer molecular dynamics (BOMD) and Car-Parrinello molecular dynamics (CPMD), so as to render amorphous structures with unprecedented accuracy and efficiency. Here we apply SGCPMD simulations to the topical black titania, and successfully reveal its precise electronic structure, offering fresh insights into the color origin and catalytic mechanisms.

The focus on black titania here is needed. Since the first report in 2011 by Chen *et al.*[4], black titania has been intensively studied. Fabricated using various methods such as hydrogenation, thermal reduction, and magnesium reduction[11-13], black titania materials are found to feature disordered surface layer, from entirely amorphous structures to core-shell (amorphous-shell and crystalline-core) architectures[14-18]. Efforts have been made to elucidate the origin of its black appearance and to improve its photocatalytic performance, but there is no consensus so far on the structure-property relationships and how to realize its full potential in photocatalysis. The experimental progress is also slow: the HER efficiency has only increased 4.3 times since the first report in 2011[11,19]. remaining far below the practical level. Meanwhile, theoretical studies mostly focused on the crystalline phases of titania containing the defects of interest (e.g., oxygen vacancies or dopants), using the density functional theory (DFT); and no precise electronic structure has been reported on amorphous titania to date.

**Results**

An AIMD-based melt melt-quench quench scheme (detailed in Supporting Information) was carried out to generate $a$-TiO$_x$ models after randomly removing some oxygen atoms from anatase titania (Figs. 1a and S1). The average mean force deviation was unbiased during the melting simulation at 3000 K (Fig. 1b), so the damped force was taken as a thermostat in our Langevin ensemble (the white noise of the force has the same features as the intrinsic noise $\Xi_D$ in Langevin equation). The kinetic energy distribution of $a$-TiO$_x$ at 3000 K was used as a criterion for a reasonable SGCPMD simulation as it fitted well with the Maxwell-Boltzmann distribution (Fig. 1c)[7]. The propagation of the electronic degrees of freedom was quite stable even at 3000 K. The radial distribution function of TiO pair (cutoff = 2.4 Å) collected from 50 ps equilibrium trajectories of $a$-TiO$_x$ indicates that the amorphous structures were successfully generated (Fig. S4).

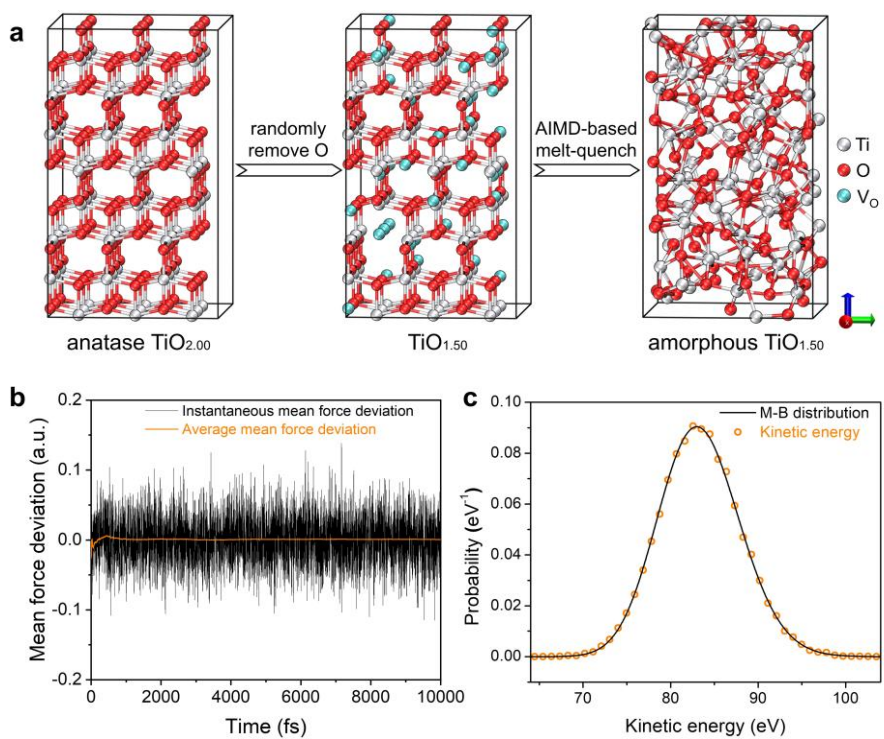

**Fig. 1 *Ab initio* molecular dynamics simulations. a**, Schematic of the *a*-TiO$_{1.50}$ modelling procedure. Silver, red and cyan spheres indicate Ti, O and oxygen vacancy, respectively. **b**, Mean force deviation of a 10-ps equilibration trajectory extracted from the melting simulation at 3000 K. The average mean force deviation is unbiased, indicating that the molecular dynamics simulation is accurate. **c**, The kinetic energy distribution fitting plot of an equilibration trajectory at 3000 K. "M-B" stands for Maxwell-Boltzmann.

With the precise atomic models of *a*-TiO$_x$ at hand, snapshots were selected from the equilibrium trajectory to examine the electronic structures using the DFT and HSE06 hybrid functional[20,21]. The projected density of states are shown in Fig. 2a. Two main features in the electronic structure stand out. First, as *x* decreases (*a*-TiO$_x$; *x* from 2.00 to 1.75, 1.50, 1.25 and 1.00), the valence band (VB) monotonically drifts away from the Fermi level, indicating that shorter-wavelength light would be needed to excite electrons from the VB into the empty states in the conduction band (CB). Second, for *a*-TiO$_x$ (x < 2), additional occupied electronic states appear in the CB tail. The electronic states in the CB and its tail all mainly originate from the Ti 3*d* orbitals. The bonding nature of the Ti-O pairs in *a*-TiO$_x$ is revealed by the crystal orbital Hamilton population (COHP)[22] and crystal orbital bond index (COBI)[23] methods (Fig. S9). The bond order of Ti-O pairs in $a$-TiO$_x$ is quantitatively evaluated via the integrated COBI (ICOBI), which indicates that the covalent bonding of Ti-O pairs in $a$-TiO$_x$ weakens as x becomes smaller (Fig. S10). The calculated absorption spectra of anatase and $a$-TiO$_x$ (Fig. 2b) feature, at a smaller x, higher absorption of visible light and thus a darker appearance. These two main features therefore clarify the blackening mechanism of $a$-TiO$_x$. Namely, the blackening is not due to the commonly believed bandgap reduction, but instead to the photoexcitation from the CB tail to the upper CB, as also agrees with the experimental observations on the localized excitation *via d-d* transitions from the mid-gap states to the empty states in reduced TiO$_2$[24]. Notably, these visible-light-excited charge carriers can easily recombine and are therefore difficult to be harnessed, which explains why the photocatalytic performance of black titania is often below expectation.

The variation of the atomic coordination environment in $a$-TiO$_x$ is informative (Fig. 2c). Statistics on the coordination numbers (CN, from 3 to 7, cutoff = 2.4 Å) of Ti-O pairs show that, when over 50% of the O atoms are removed, the fraction of 6-coordinated Ti plummets while the fractions of 3- and 4-coordinated Ti rise. The counts of Ti-Ti pairs in $a$-TiO$_x$ (cutoff = 3.1 Å) are inversely proportional to x in $a$-TiO$_x$, indicating stronger overlapping of Ti 3$d$ orbitals at a smaller x. Further, unlike crystalline titania where a vacancy is created upon removal of an oxygen atom from the lattice, in $a$-TiO$_x$, oxygen vacancy and low-valence Ti (e.g., Ti$^{3+}$) are less distinct: namely, oxygen deficiency is not localized but "spread out" through amorphization (i.e., a clear-cut vacancy is hardly spotted). The wavefunctions of electronic states at the CB minimum (CBM) in real-space (Fig. 2d) show that the Anderson localization in $a$-TiO$_{2.00}$ ($E$-$E_f$ = 0.17 eV) features the non-overlapped Ti 3$d$ orbitals. As the reduction level rises, the localization at CBM turn to the overlapped Ti 3$d$ orbitals and metal clusters are formed, e.g., the Ti$_4$ tetrahedron in $a$-TiO$_{1.50}$ and $a$-TiO$_{1.25}$ and the Ti$_8$ metal-like cluster in $a$-TiO$_{1.00}$. It is further noted that, for $a$-TiO$_x$ (x ≤ 1.25), more than one Ti clusters are spotted in the electronic state. Thus, the localized states become extended as the reduction progresses. The emergence of metal clusters indicates a "disproportionation-like" process in $a$-TiO$_x$, suggesting a distinct structural response to "consume" the oxygen deficiency; namely, the presence of low-valence Ti does not necessitate oxygen vacancy, which, incidentally, also helps settle the long-standing debate on the coexistence of oxygen vacancy/Ti$^{3+}$ in reduced titania[11-13].

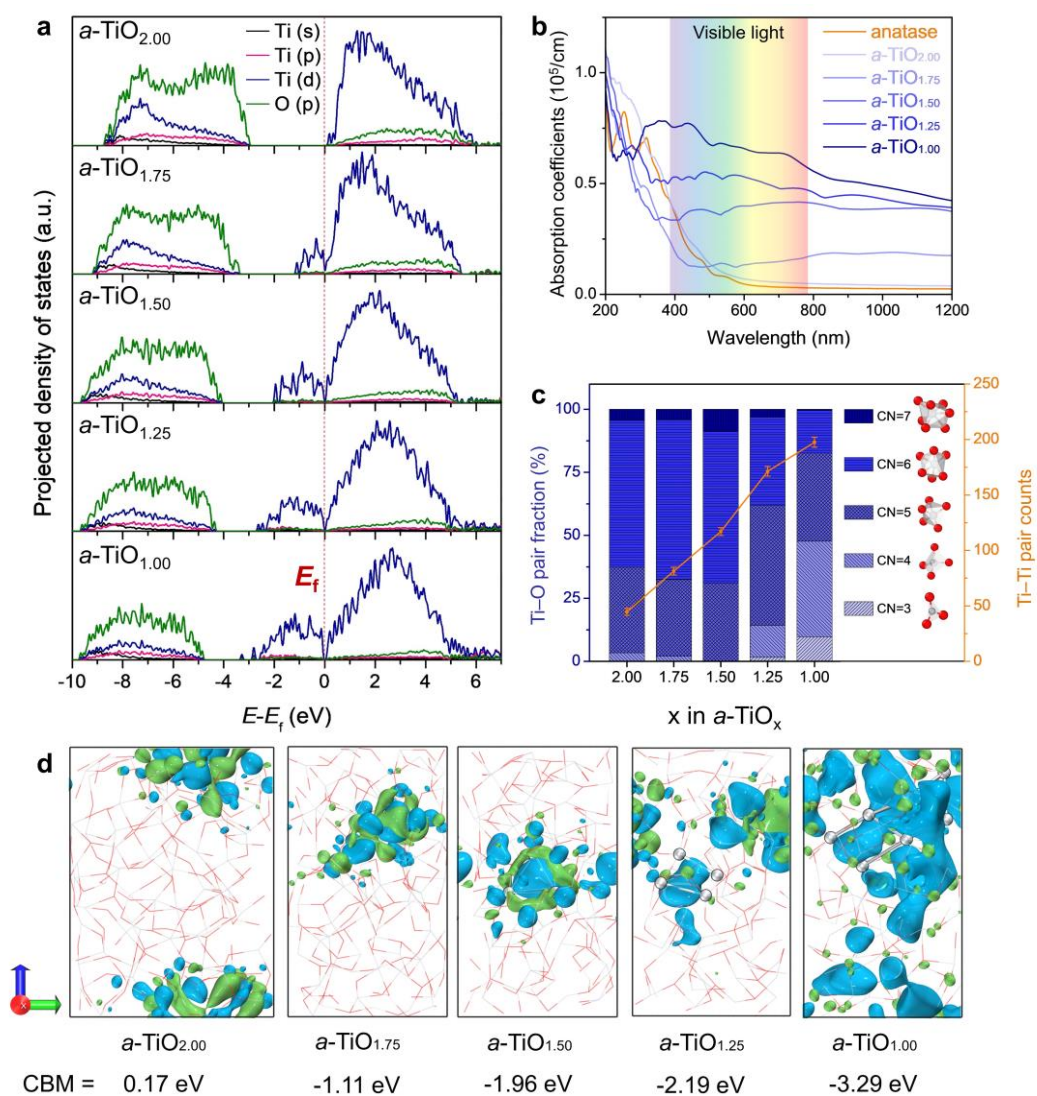

**Fig. 2 Electronic structure analysis. a**, The projected density of states of $a$-TiO$_x$ (x = 2.00, 1.75, 1.50,

1.25 and 1.00). **b**, Calculated optical absorption coefficients of anatase and *a*-TiO$_x$. **c**, Left axis: the statistical histogram of the Ti-O pairs with different coordination numbers (3-7, cutoff = 2.4 Å) in *a*-TiO$_x$. The corresponding atomic models are shown in the legends. Right axis: the counts of Ti-Ti pairs in *a*-TiO$_x$ (cutoff = 3.1 Å). **d**, The wavefunctions of electronic states at the conduction band minimum of *a*-TiO$_x$ in real space. Cyan and lime iso-surfaces represent the positive and negative phases of wavefunctions, respectively. Isovalue = 1.36 × 10$^{-7}$.

In 1958, the absence of diffusion of the mobile entities in a random lattice, so-called "Anderson localization", was proposed (*25*). The idea of "mobility edge" that separates the localized and the extended states came out later in 1967[26,27]. Here we calculate the inverse participation ratios (IPR) of the electronic states in *a*-TiO$_x$ (Fig. 3a) to quantitatively characterize the degree of localization of the Kohn-Sham eigenvalues[28]. The IPR value is related to the local density of states *via* IPR = $\sum_i |\Psi_{\alpha,i}|^4 / (\sum_i |\Psi_{\alpha,i}|^2)^2$, where $\Psi_{\alpha,i}$ is projection of the Kohn-Sham orbital on each atom. The IPR values are low (approximately 1/N, where N is the total number of atoms in the model) for the extended states but significantly higher for the localized states. From the IPR values of *a*-TiO$_x$, the "mobility edges" ($E_V$ for VB and $E_C$ for CB, Fig. 3b) dividing the extended and localized electronic states are revealed.

For crystalline TiO$_2$, all the electronic states are extended, according to the band theory (Fig. S7). Upon amorphization (i.e., *a*-TiO$_{2.00}$), the upper and lower edges of both the VB and CB become localized (Anderson localization) (Figs. 3a,b). After *a*-TiO$_2$ is reduced (i.e., x < 2), new localized states appear beneath the CB tail. Interestingly, as reduction progresses, these localized states become increasingly extended (Fig. S8), indicating a filling-controlled insulator-metal Mott-Hubbard transition (IMT)[29,30], i.e., stronger overlap of the Ti *3d* orbitals occurs at the same energy level. Note that optical bandgap of $a$-TiO$_x$ is larger than the gap between the VB maximum (VBM) and the CBM, due to the Anderson localization at the VBM and CB and the fact that the electronic states at the CB tail are all occupied (Fig. 3c). If the localized electrons at the CB tail (induced by reduction) are removed by heterojunctions (with metals or semiconductors) or electrical bias, $a$-TiO$_x$ can utilize visible light for photoexcitation from the VBM (mainly the O 2p orbitals) to CBM (mainly the Ti 3d orbitals) and the electron-hole pairs thus produced do not easily recombine (Fig. 3c). This also explains why the reported best-performing black titania-based photocatalysts often feature heterojunctions; for example, Pt as a co-catalyst can boost the HER efficiency by several orders of magnitude[19,31].

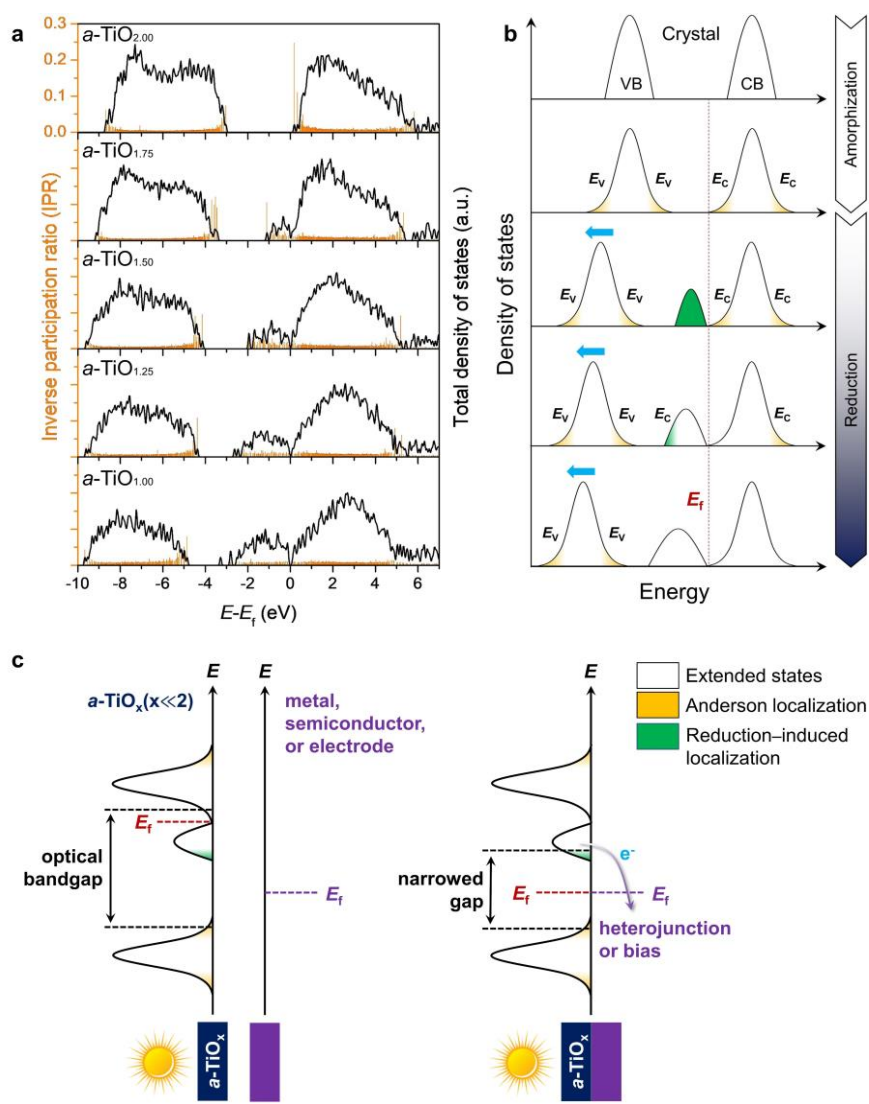

**Fig. 3 Schematic diagram of the catalytic mechanisms. a**, The inverse participation ratio (IPR) values of the electronic states and the total density of states for $a$-TiO$_x$. **b**, Generic evolution patterns of the electronic structures of covalent compounds undergoing amorphization and reduction; $E_V$ and $E_C$ denote the mobility edges of the valence and conduction bands, respectively. The blue arrow indicates the energy shift of the valence band upon reduction. **c**, Schematic diagrams of the reduced amorphous titania under electrical bias or in contact with another material so as to drain the electrons in the CB tail. The electronic structure shown here is of a high reduction level ($a$-TiO$_x$, x≪2).

Single-atom-dispersed amorphous materials have attracted widespread attention for their superior catalytic performance[32]. However, it remains difficult for conventional simulation methods to elucidate the interactions between the single atoms and the amorphous host[1]. Here, we simulate the Pt single-atom-adsorbed $a$-TiO$_{2.00}$ (Fig. 4a, detailed in Supporting Information), and this is the first time surface adsorption on amorphous substances is simulated. $a$-TiO$_x$ (x = 2.00) is here selected, as a reduced state (x < 2) is found (see above) to adversely affects photocatalysis. Especially noted of the surface of $a$-TiO$_{2.00}$ is a local, open feature that readily accommodates the Pt atoms (Movie 1; while such accommodation would be impossible for the crystalline counterpart owing to the compact structure; Fig. 4b). Specifically, the adsorbed Pt single-atom bonds to a hydroxyl group while the adjacent surface Ti atom adsorbs a water molecule (Movie 2)[33]. The pDOS of the interface model reveals that the electronic states of the environmental water and the bonded hydroxyl group reside at lower energy levels, reflecting their high stability, whereas the outer orbitals of Pt ($5d$ and $6s$) hybridize and form frontier orbitals above the VBM of $a$-TiO$_{2.00}$ (Figs. 4c and S11). The electronic states of Pt thus lift the Fermi level and considerably narrow the bandgap, facilitating photoexcitation. Anderson localization at the CBM may adversely affect the photoexcitation process because a spatial mismatch may occur between the electronic states of Pt and the localized states; thus, only electronic excitation from Pt to the extended states in the CB is allowed, which means only shorter-wavelength light can be harvested.

We are now in a position to propose the photocatalytic HER process (Fig. 4d): I) At the interface of the platinized $a$-TiO$_{2.00}$ and water, Pt is anchored onto the surface Ti and O atoms and bonds to a hydroxyl group, while the adjacent surface Ti atom adsorbs a water molecule. Photoexcitation under solar irradiation generates holes in the frontier orbitals of Pt and electrons in the extended states of $a$-TiO$_{2.00}$. II) The Pt-OH bond is cleaved upon hole capture by the hydroxyl group, releasing an •OH free radical which is subsequently consumed by a sacrificial agent (e.g., methyl alcohol), whereas the photogenerated electrons diffuse through the extended states in the adjacent $a$-TiO$_{2.00}$ and reduces the water molecule adsorbed nearby, resulting in the reactive intermediate Pt-H$_{ad}$ (the Volmer step) which subsequently produces H$_2$ following the Heyrovsky step[34]. III) The consumed species are replenished from the environmental water.

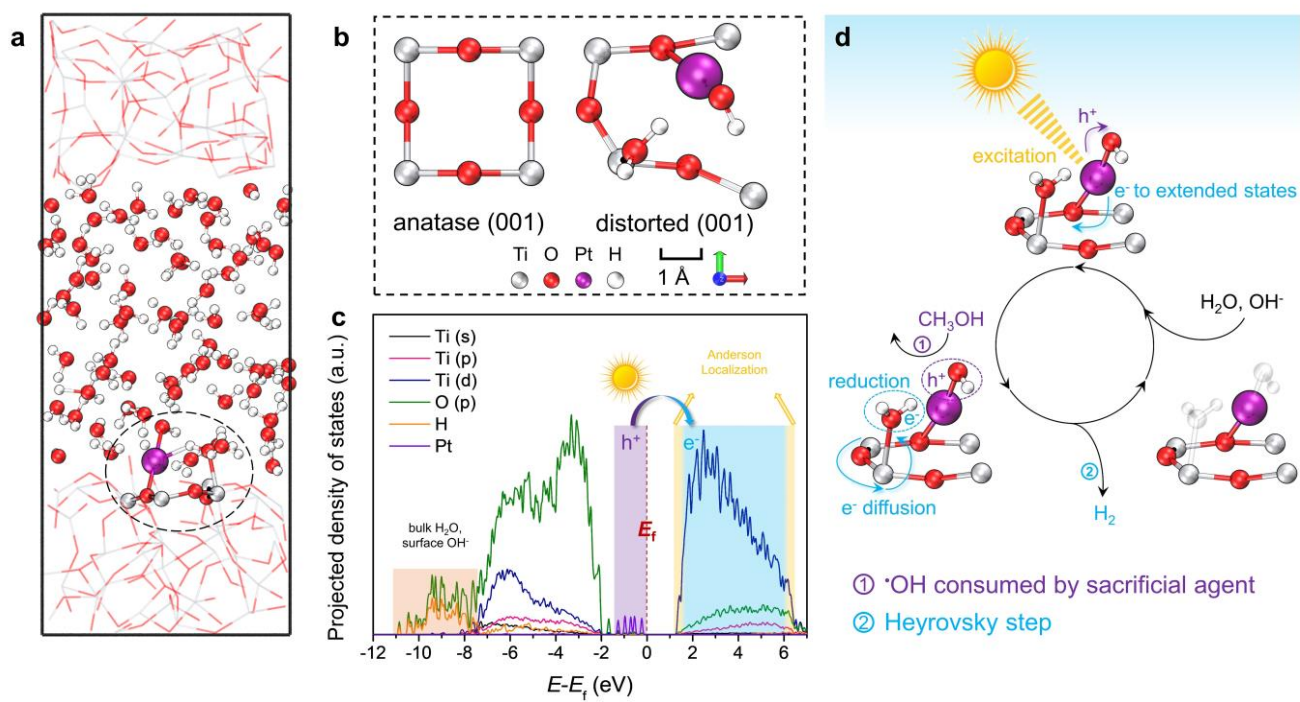

**Fig. 4 Structure-activity relationships in black titania. a**, A snapshot of the simulated model of the interface between Pt-adsorbed *a*-TiO$_{2.00}$ and H$_2$O. *a*-TiO$_{2.00}$ and water molecules are denoted with lines and a ball-and-stick model, respectively. The dashed circle highlights the adsorption of a Pt atom onto the *a*-TiO$_{2.00}$ surface. **b**, Anatase (001) facet and the distorted TiO$_2$ lattices at the surface of *a*-TiO$_{2.00}$. **c**, The projected density of states of the model of the interface between Pt-adsorbed *a*-TiO$_{2.00}$ and H$_2$O. **d**, Proposed mechanism of the hydrogen evolution reaction process over Pt-adsorbed *a*-TiO$_{2.00}$. The hole (h$^+$) generated upon photoexcitation is received by the hydroxyl adsorbed on a Pt atom and is eventually consumed by a sacrificial agent (e.g., methyl alcohol). The excited electron (e$^-$) diffuses through the extended states and is eventually consumed through the reduction of the water molecule adsorbed on the adjacent Ti atom.

The underlying mechanism elucidated in this work offers guidance in designing a broader array of new systems of black titania for better catalytic performance. For example, *a*-TiO$_x$ with a high reduction level features minimal localized states at the CB tail and a narrow gap (e.g., 1.57 eV) between VBM and CBM (Figs. 3a and S6). This material is thus very promising for efficient visible-light photocatalysis, if its $E_f$ can be lowered under the CB tail. To achieve this goal, a possible solution is to empty the occupied electronic states in the CB tail through heterojunction structures (with crystalline titania, metals, or other semiconductors) or electrical bias[19,31].

**Conclusion**

In summary, the precise electronic structure of amorphous black titania is calculated through the SGCPMD simulations with the DFT and HSE06 hybrid functionals. It is discovered that the black color arises from the reduction of amorphous titania, with the resulted conduction band tail being detrimental to photocatalytic properties (i.e., the black color does not help photocatalysis). The generic evolution patterns of the electronic structures for covalent compounds undergoing amorphization and reduction were also revealed: as the degree of reduction progresses, Anderson localization, Reduction-induced localization and the extended states sequentially dominate the electronic states in the CB tail. Furthermore, the amorphization of $TiO_2$ enables easy surface adsorption of Pt atoms and facilitates charge separation and transfer, which is the key factor that enhances the photocatalytic HER performance. The discoveries presented here are useful for acquiring mechanistic understandings of the unusual properties for various other amorphous materials.

**Methods**

*Ab initio* simulations and the DFT calculations are detailedly discribed in supplementary information section.

**Data Availability**

The data that support the findings of this study are available from the corresponding author, Dr.

**Acknowledgements**

This work was jointly supported by the National Key R&D Program of China (Project No. 2017YFA0204403), Shenzhen-Hong Kong Science and Technology Innovation Cooperation Zone Shenzhen Park Project (HZQB-KCZYB-2020030), the Major Program of National Natural Science Foundation of China (Project 51777178), and the Innovation and Technology Commission of HKSAR through the Hong Kong Branch of National Precious Metals Material Engineering Research Centre, the City University of Hong Kong (Project 9667207).

**Author Contribution**

JL and YYL formulated the question, initiated the work, and supervised the research. JZ performed simulations of the electronic structures. JZ and YYL plotted the figures, analyzed the data, and wrote the manuscript. ZX and JL revised the manuscript.

**Competing Interests**

The authors declare that there are no competing interests.

**Supplementary Information** for

Precise electronic structures of amorphous solids: unravelling the color origin and photocatalysis of black titania

Jing Zhong [1,2,3], Zhengtao Xu [5], Jian Lu [1,3,4,6,*], Yang Yang Li [1,2,3,4,6,*]

[1] Hong Kong Branch of National Precious Metals Material Engineering Research Centre, City University of Hong Kong, Hong Kong, China

[2] Center of Super-Diamond and Advanced Films (COSDAF), City University of Hong Kong, Hong Kong, China

[3] Department of Materials Science and Engineering, City University of Hong Kong, Hong Kong, China

[4] Department of Mechanical Engineering, Greater Bay Joint Division, Shenyang National Laboratory for Materials Science, City University of Hong Kong, Hong Kong, China

[5] Department of Chemistry, City University of Hong Kong, Hong Kong, China

[6] Centre for Advanced Structural Materials, City University of Hong Kong Shenzhen Research Institute, Greater Bay Joint Division, Shenyang National Laboratory for Materials Science, 8 Yuexing 1st Road, Shenzhen Hi-Tech Industrial Park, Nanshan District, Shenzhen, China

* Email: jianlu@cityu.edu.hk; yangli@cityu.edu.hk

**Simulation details**

We perform second-generation Car–Parrinello molecular dynamics simulations[1] in the open-source CP2K package[2, 3] with the canonical ensemble, using Langevin dynamics to propagate the equations of motion (eqs. 1 and 2):

$$M_I \ddot{R}_I = F_{BO} - (\gamma_D + \gamma_L)\dot{R}_I + \Xi_I \quad (1)$$

$$M_I \ddot{R}_I = F_{ASPC} - \gamma_L \dot{R}_I + \Xi_I \quad (2)$$

where $M_I$ is the ionic mass, $F_{ASPC}$ is the force calculated through always stable predictor–corrector (ASPC) extrapolation[4] as implemented in CP2K, and $\Xi_I$ is a random noise defined by eq. 3:

$$\Xi_I = \Xi_D + \Xi_L \quad (3)$$

$\Xi_I$ must obey the fluctuation and dissipation theorem to ensure accurate sampling of the Maxwell–Boltzmann distribution (eq. 4):

$$\langle \Xi_I(0)\Xi_I(t) \rangle = 6(\gamma_D + \gamma_L)M_I k_B T \delta(t) \quad (4)$$

where $\gamma_L$ is a Langevin friction coefficient, and $\gamma_D$ is the intrinsic friction coefficient corresponding to the random noise $\Xi_D$. In this work, $\gamma_L$ is set to 0.2 fs$^{-1}$ throughout the simulations; $\gamma_D$ is varied until the value that satisfies the equipartition theorem $\langle \frac{1}{2}M_I \dot{R}_I^2 \rangle = \frac{3}{2}k_B T$ is obtained. The $\gamma_D$ value is temperature-dependent because it controls the dissipation of energies (potential and kinetic) in MD simulations. $\gamma_D$ is set at 1.5 × 10$^{-3}$ and 1.0 × 10$^{-3}$ fs$^{-1}$ for TiO$_x$ systems at 3000 and 300 K, respectively. It also varies among atomic systems. For H$_2$O and Pt, $\gamma_D$ is set at 5 × 10$^{-5}$ and 2.2 × 10$^{-4}$ fs$^{-1}$, respectively.[5]

The electronic structure is described using a mixed Gaussian and plane waves[6] approach and the density functional theory (DFT). The Kohn–Sham orbitals are expanded in a Gaussian-type basis set of triple-zeta valence plus polarisation[7] quality, while the interaction between the cores is described by Goedecker–Teter–Hutter pseudopotentials and Perdew–Burke–Ernzerhof functionals.[8-10] The cutoff is set at 300 Ry. Periodic boundary conditions are used throughout the simulations. The time step is set at 1 fs. The Brillouin zone is integrated only at the Γ point of the models.

The amorphous models of reduced titania ($a$-TiO$_x$, $x$ = 2.00, 1.75, 1.50, 1.25 and 1.00) were generated following a melt–quench approach. Starting from a 3 × 3 × 2 supercell (containing 216 atoms) of anatase, we randomly removed certain numbers of oxygen atoms (i.e., 0, 18, 36, 54 and 72 oxygen atoms for $x$ = 2.00, 1.75, 1.50, 1.25 and 1.00, respectively) from the anatase model to represent different 'levels' of reduction. The resultant TiO$_x$ models were equilibrated at 3000 K for 30 ps to erase the crystalline structural memory. The randomised structure was then quenched down to 300 K for over ~60 ps (Fig. S2). Note that for TiO$_{2.00}$, we adjusted the initial positions of atoms to facilitate the melting of anatase, which would not affect our results since the positions of atoms would be fully randomized during the equilibration at 3000 K. The structure was equilibrated at 300 K for a further 50 ps, while the volume was adjusted to keep the pressure approaching zero (Fig. S3). The cutoffs for Ti–O and Ti–Ti pairs were set at 2.4 Å and 3.1 Å, respectively. We modelled the interface between Pt-adsorbed $a$-TiO$_{2.00}$ and H$_2$O by adding 72 water molecules to the vacuum space between the bulk surface of $a$-TiO$_{2.00}$. The Pt atom was first randomly placed on the surface of $a$-TiO$_{2.00}$, and the model was equilibrated at 600 K for 50 ps. The Ti atoms near the Pt atom adjusted their positions for better interaction with the Pt atom until equilibrium was reached (Movie. 1 and Fig. S12). We set the time step at 0.5 fs, considering the vibration frequency of the water molecules. The hydroxylation process on the adsorbed Pt atom was extracted from the equilibration trajectory (Movie. 2).

DFT calculations were performed using projector augmented wave pseudopotentials[12] and Perdew–Burke–Ernzerhof functionals[10] in the Vienna *ab initio* Simulation Package (VASP).[11] The energy cutoff of plane waves was set at 500 eV. Only the Γ point was sampled in the Brillouin zone. The HSE06 exchange–correlation hybrid functional[13] was used to calculate the electronic structure. The self-consistent field energy convergence was set at $10^{-7}$ eV. The VASPKIT code was used for data post-processing.[14] Quantum chemistry bonding analyses were performed using the schemes of crystal orbital Hamilton population and crystal orbital bond index,[15, 16] implemented in the Local Orbital Basis Suite Towards Electronic-Structure Reconstruction (LOBSTER) code.[17]

The optical absorption spectra were calculated using the HSE06 functional. The imaginary part of the dielectric function $\varepsilon_2$ was calculated using a previously reported equation[18] that requires the conduction and valence band states as input. The absorption coefficient $\alpha(\omega)$ was calculated using eq. 5:

$$\alpha(\omega) = \sqrt{2}\omega\left(\sqrt{\varepsilon_1^2(\omega) + \varepsilon_2^2(\omega)} - \varepsilon_1(\omega)\right)^{1/2} \quad (5)$$

where $\varepsilon_1$ is the real part of the dielectric function; it was obtained from $\varepsilon_2$ using the Kramer–Kronig relationship.

The charge density difference was obtained from eq. 6:

$$\Delta\rho = \rho_{(Pt@a\text{-}TiO_2+water)} - \rho_{Pt} - \rho_{(a\text{-}TiO_2+water)} \quad (6)$$

where $\rho_{(Pt@a\text{-}TiO_2+water)}$ is the charge density of the total adsorption system, $\rho_{Pt}$ is the charge density of the Pt atom and $\rho_{(a\text{-}TiO_2+water)}$ is the charge density of the model of the *a*-TiO$_2$/water interface.

Table S1. Lattice parameters of simulated models of $a$-TiO$_x$.

| Models | x (Å) | y (Å) | z (Å) |
| --- | --- | --- | --- |
| $a$-TiO$_{2.00}$ | 12.53 | 12.53 | 16.71 |
| $a$-TiO$_{1.75}$ | 11.02 | 11.02 | 18.83 |
| $a$-TiO$_{1.50}$ | 10.50 | 10.50 | 17.94 |
| $a$-TiO$_{1.25}$ | 10.16 | 10.16 | 17.36 |
| $a$-TiO$_{1.00}$ | 9.86 | 9.86 | 16.85 |

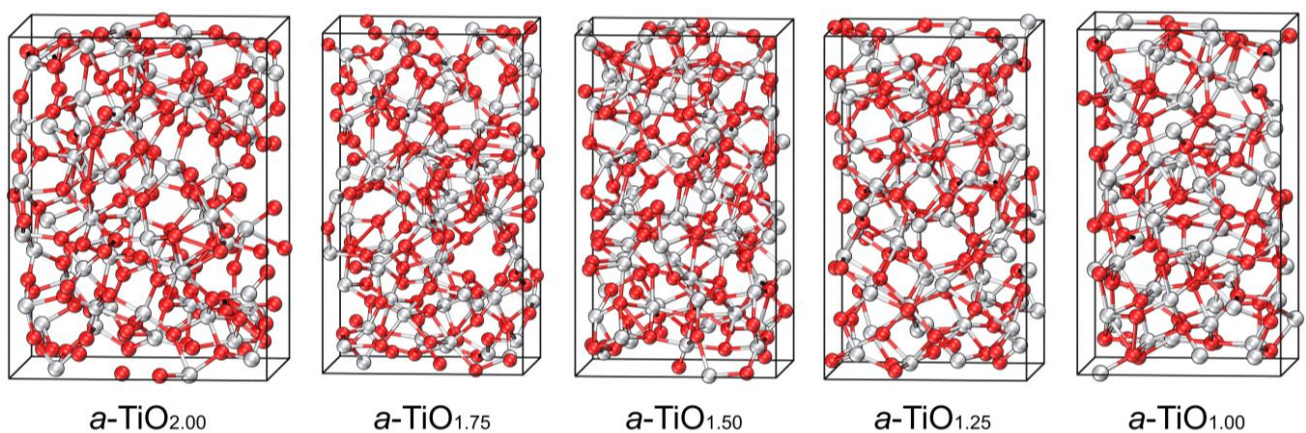

**Fig. S1** | Snapshots of $a$-TiO$_x$ during equilibration at 300 K.

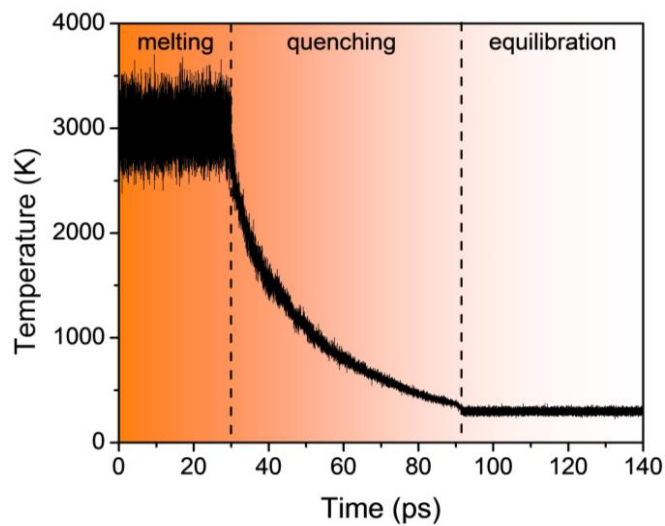

**Fig. S2 |** Temperature profile as a function of simulation time during an *ab initio* molecular dynamics run.

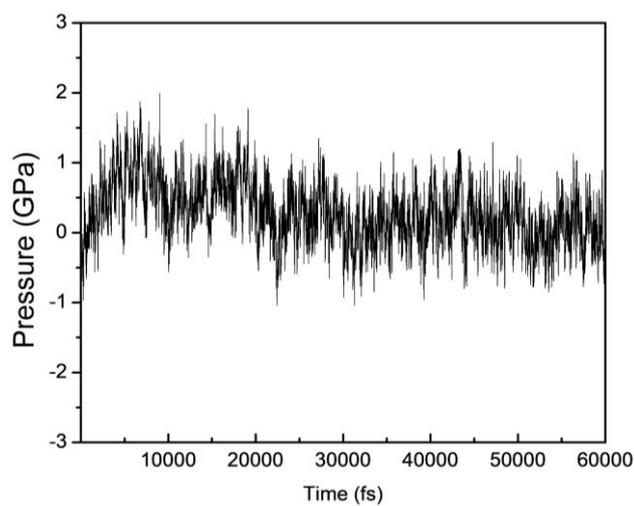

**Fig. S3 |** The pressure profile of *a*-TiO$_{1.50}$ as a function of simulation time during equilibration at 300 K; the profile demonstrates our adjustment to keep the pressure approaching zero.

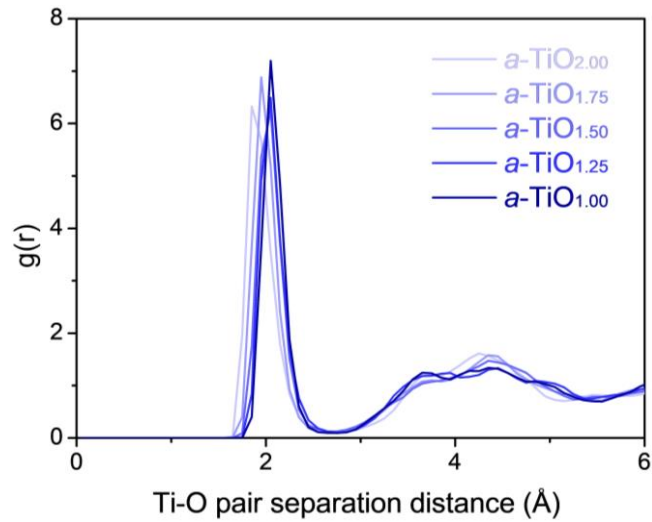

**Fig S4** | The radial distribution function of the Ti–O pair (cutoff = 2.4 Å) collected from 50-ps equilibration trajectories of $a$-TiO$_x$. The first-layer distance of the Ti–O pair becomes slightly larger as the value of $x$ in $a$-TiO$_x$ is reduced.

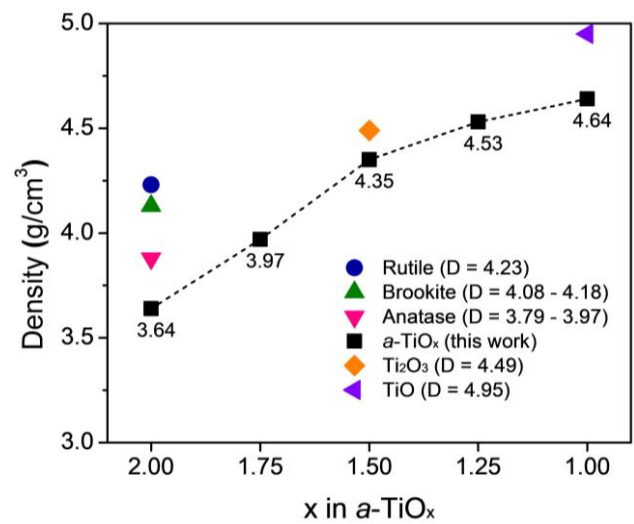

**Fig. S5** | Densities of $a$-TiO$_x$ simulated in this study and crystalline TiO$_x$.[19]

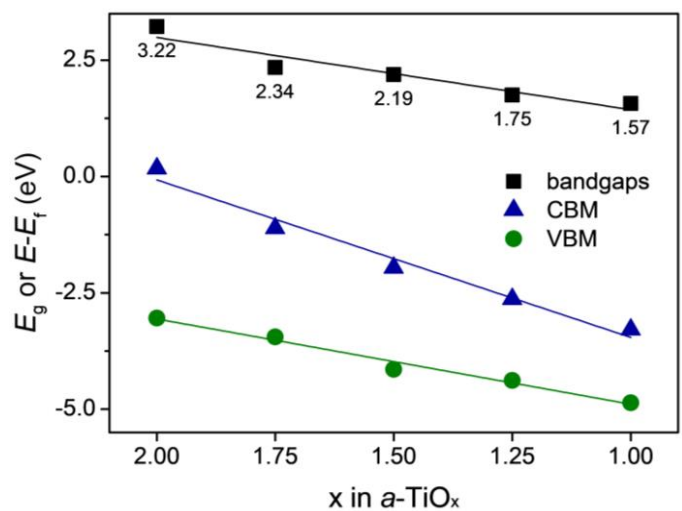

**Fig. S6 |** Fitting lines of the bandgaps (black square), conduction band minimum (CBM, blue triangle) and valence band maximum (VBM, green circle) for $a$-TiO$_x$.

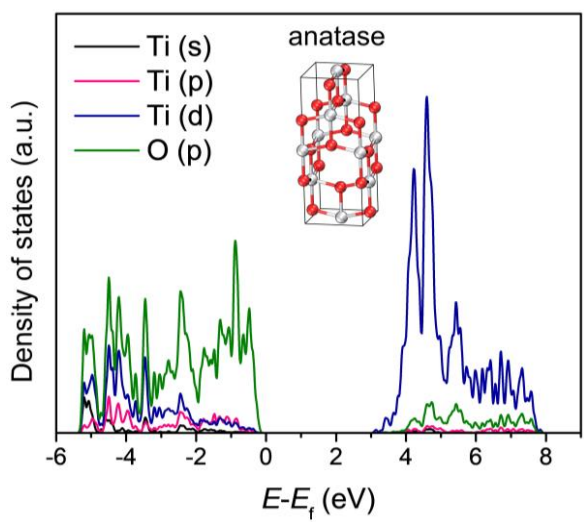

**Fig. S7 |** Electronic density of states for anatase. A 7 × 7 × 3 Monkhorst–Pack k point mesh was used to sample the unit cell.

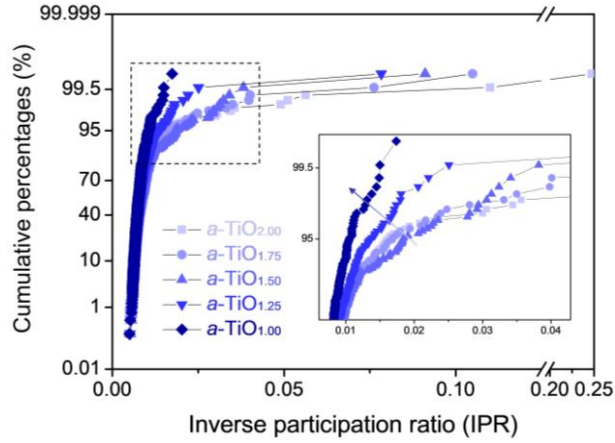

**Fig. S8 |** The cumulative percentages of IPR at the CB (from a CB tail greater than $E_f$ by up to 6 eV) of *a*-TiO$_x$. Inset: the magnified image of a selected region; the arrow indicates that as *x* in *a*-TiO$_x$ decreases, the IPR decreases.

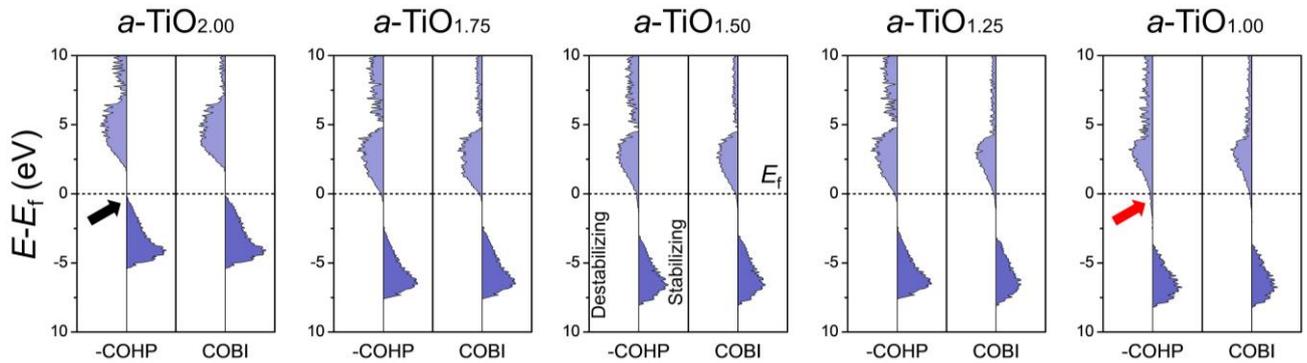

**Fig. S9 |** The crystal orbital Hamilton population (COHP) and crystal orbital bond index (COBI) of *a*-TiO$_x$.

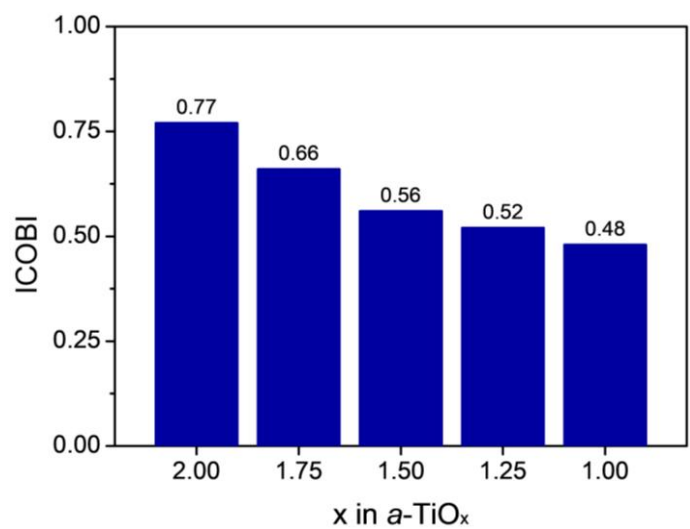

**Fig. S10** | The integrated crystal orbital bond index (ICOBI) of $a$-TiO$_x$.

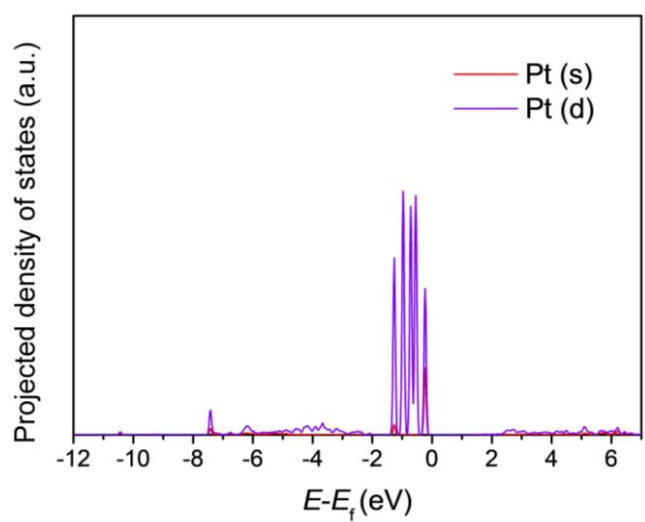

**Fig. S11** | The projected density of states for adsorbed Pt 6$s$ and 5$d$ orbitals.

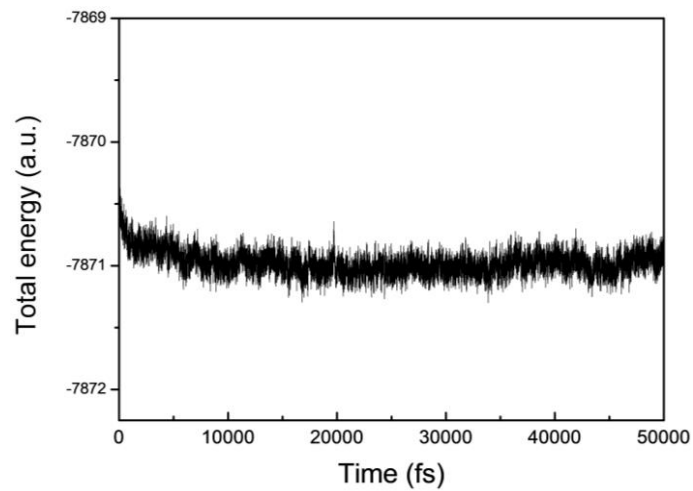

**Fig. S12** | The total energy profile of the model of the interface between Pt-adsorbed $a$-TiO$_{2.00}$ and H$_2$O as a function of simulation time during equilibration at 600 K.

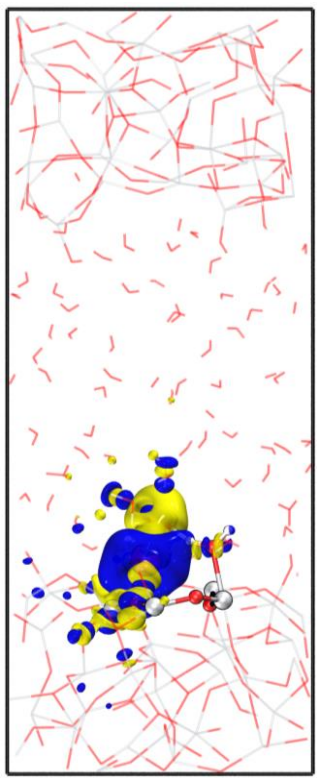

**Fig. S13** | The charge density differences for the interface of platinised $a$-TiO$_{2.00}$ and water. Blue and yellow iso-surfaces represent depletion and accumulation, respectively. Isovalue = 0.002 e/Å$^3$. The distorted (001) surface is shown as a ball-and-stick model, while H$_2$O molecules and $a$-TiO$_{2.00}$ are shown as lines.